\documentclass[aps,prd,preprint,amsmath,amssymb]{revtex4}
\usepackage{epsfig,float}
\usepackage{graphicx}
\usepackage{dcolumn}
\usepackage{bm}

\newcommand{\be}{\begin{equation*}}
\newcommand{\ee}{\end{equation*}}
\newcommand{\bea}{\begin{eqnarray}}
\newcommand{\eea}{\end{eqnarray}}
\newcommand{\bean}{\begin{eqnarray*}}
\newcommand{\eean}{\end{eqnarray*}}

\begin{document}
\title{Nucleon Resonances with Hidden Charm in Coupled-Channel Models}

\author{Jia-Jun Wu$^{1,2}$, T.-S. H. Lee$^{2}$ and B.~S.~Zou$^{1,3}$ \\
$^1$ Institute of High Energy Physics, CAS, P.O.Box 918(4), Beijing 100049, China\\
$^2$ Physics Division, Argonne National Laboratory, Argonne, Illinois 60439, USA\\
$^3$ Theoretical Physics Center for Science Facilities, CAS,
Beijing 100049, China}

\date{Feb. 4, 2012}

\begin{abstract}
The model dependence of the predictions of nucleon resonances
with hidden charm is
investigated.
We consider several coupled-channel models which are derived from relativistic quantum
field theory by using (1) a unitary transformation method, and (2)
 the three-dimensional reductions of Bethe-Salpeter Equation.
With the same vector meson exchange mechanism,
we find that all models give very narrow molecular-like nucleon resonances with
hidden charm in the mass range of 4.3 GeV $ < M_R < $ 4.5 GeV,
in consistent with the previous predictions.

\end{abstract}


\maketitle

\section{Introduction}

In the classical quark models, each baryon is made of three
constituent quarks~\cite{PDG}. The pattern of the spectra and the
static properties of the ground and low-lying excited states of
baryons can be described reasonably well within these models.
However, there are large deviations between the predictions from
these models and the experimental data~\cite{capstick}, such as the
strong coupling of $N^*(1535)$ to the strangeness, and the mass
order between $N^*(1535)$ and $\Lambda^*(1405)$. In the classical
3-quark models, the $N^*(1535)$ with $(uud)$-quarks is expected to
be lighter than $\Lambda^*(1405)$ with $(uds)$-quarks. This problem
may be solved by the penta-quark picture for these excited baryons.
In the penta-quark models, the $N^*(1535)$ with $[uu][ds]\bar{s}$ is
naturally heavier than the $\Lambda^*(1405)$ with
$[ud][sq]\bar{q}$~\cite{Liubc}. Actually, the conventional orbital
excitation energy of a original constituent quark in a baryon is
already comparable to drag out a $q\bar q$ pair from the gluon
field. As a result, some excited baryons are proposed to be
meson-baryon dynamically generated states~\cite{Weise,or,Osetoller,
meiss,Inoue, lutz,Hyodopk} or states with large ($qqqq\bar q$)
components~\cite{Riska,Liubc,Zou10}. But because of the same
 resonances predicted by  different models are in the similar energy region,
there are always some adjustable ingredients in each model to fit
the experimental data. Thus it is difficult to pin down the nature
of these baryon resonances. One way to avoid such difficulty is to
replace light flavor $q\bar q$ in these baryons by $c\bar{c}$.
Brodsky $et\ al.$~\cite{brodsky} proposed in the early 1980s that
there are about $1\%$ $uudc\bar{c}$ components in the proton.
Recently, Refs.\cite{oset,wang,liuxiang} have used different methods
to predict some narrow hidden charm $N^*_{c\bar{c}}$ and
$\Lambda^*_{c\bar{c}}$ resonances with masses above 4 GeV and widths
smaller than 100 MeV. These resonances, if observed, absolutely
cannot be ascribed to the conventional 3-quark states. Therefore, it
is important to investigate the extent to which the predicted
$N^*_{c\bar{c}}$ and $\Lambda^*_{c\bar{c}}$ resonances can be
further firmly established.

In this work we focus on the predictions~\cite{oset} from a
Beijing-Valencia collaboration. Their results are from solving the
following algebraic coupled-channel equations
\begin{eqnarray}
T_{\alpha,\beta}(s) =
\sum_{\gamma}V_{\alpha,\gamma}(s)\hat{G}_\gamma(s)T_{\gamma,\beta}(s)+V_{\alpha,\beta}(s)
\label{eq:vi-eq}
\end{eqnarray}
where $\alpha,\beta,\gamma = \Bar{D}\Sigma_c, \bar{D}\Lambda_c,
\eta_c N$,  and $s$ is the square of the C.M. energy. In
Eq.(\ref{eq:vi-eq}) the meson-baryon potential is based on the
vector meson-exchange mechanisms of Ref.\cite{or} and is written in
the following separable form
\begin{eqnarray}
V_{\alpha,\alpha}(s)&=&\frac{C_{\alpha,\alpha}}{4f^2}(2E_{M_{\alpha}}),\nonumber\\
V_{\alpha,\beta}(s)&=&-C_{\alpha,\beta}\frac{m^2_{\rho}}{4f^2}\frac{E_{M_{\alpha}}
+E_{M_{\beta}}}{m^{2}_{M_{\alpha}}+m^{2}_{M_{\beta}}-2E_{M_{\alpha}}E_{M_{\beta}}-m^{2}_{V}}
~~~~~~(\alpha\neq\beta) \label{eq:vi-v}
\end{eqnarray}
where the $E_{M_{\alpha}}$ is the on-shell energy of the $\alpha$ channel's meson,
and $m_V$ is the mass of exchange vector. The factorized propagator $\hat{G}_\gamma(E)$
is calculated from  either using  the dimensional regularization or introducing a
cutoff parameter $\Lambda$
\begin{eqnarray}
\hat{G}(E) \rightarrow G_{DR}(E)&=&\frac{2m_{B}}{16\pi^2}\big\{a_{\mu}
+\textmd{ln}\frac{m^{2}_{B}}{\mu^{2}}+\frac{m^{2}_{M}-m^{2}_{B}+s}{2s}\textmd{ln}\frac{{m^{2}_M}}{m^{2}_{B}}\nonumber\\
&+&\frac{\bar{q}}{\sqrt{s}}\big[\textmd{ln}(s-(m^{2}_{B}-m^{2}_{M})+2\bar{q}\sqrt{s})+\textmd{ln}(s+(m^{2}_{B}-m^{2}_{P})+2\bar{q}\sqrt{s})\nonumber\\
&-&\textmd{ln}(-s-(m^{2}_{B}-m^{2}_{M})+2\bar{q}\sqrt{s})-\textmd{ln}(-s+(m^{2}_{B}-m^{2}_{M})+2\bar{q}\sqrt{s})\big]\big\}\   \label{eq:vi-g1} \\
\hat{G}(E) \rightarrow G_{C}(E)&=& \int^{\Lambda}_{0}\frac{q^{2}dq}{4\pi^{2}}
\frac{2m_{B}(\omega_{M}+\omega_{B})}{\omega_{M}\,\omega_{B}\,(s-(\omega_{M}+\omega_{B})^{2}+i\epsilon)}\  \label{eq:vi-g2}
\end{eqnarray}
where $\bar{q}$ is on shell three momentum of $MB$ system, $\omega_{M}=\sqrt{q^{2}+m^{2}_{M}}$ and
$\omega_{B}=\sqrt{q^{2}+m^{2}_{B}}$. It was found that the solutions
of the above equations yield few narrow resonances above 4.0 GeV.

As discussed in Ref.\cite{oset}, Eqs.(\ref{eq:vi-eq})-(\ref{eq:vi-g2})
are derived from making
approximations on the  Bethe-Salpeter (BS) equation. Schematically,
the BS equation  is (omitting the channel indices)
\begin{eqnarray}
T(q',q,P) &=& V(q',q,P) \nonumber \\
&&+ \int d^4kV(q',k,P)\frac{1}{((\frac{P}{2}+k)^2-m^2_M+i\epsilon)
((\frac{P}{2}-k)^2-m^2_B)+i\epsilon}T(k,q;P) \label{eq:bs-eq}
\end{eqnarray}
where $P$ is the total four momentum of the system, $q, q'$ and $k$
are  the relative momenta. For the considered vector meson-exchange,
the interaction kernel is
\begin{eqnarray}
V(q,k,P) &=& C\frac{1}{(q-k)^2-m^2_V} \label{eq:bs-k}
\end{eqnarray}
where $C$ is a coupling constant. The complications in solving
Eq.(\ref{eq:bs-eq}) is well known, as discussed in, for example,
Ref.\cite{afnan} for $\pi N$ scattering. Thus  approximations, such
as those used~\cite{oset} in obtaining
Eqs.(\ref{eq:vi-eq})-(\ref{eq:vi-g2}),
 are needed for practical calculations. There exist other approximations to
solve BS equations and alternative approaches to derive
 practical hadron reaction models
from relativistic quantum field theory.
It is thus necessary to investigate the extent to which
 the results from Ref.\cite{oset}
depend on the approximations employed. This is the  objective of
this work. We will consider several coupled-channel models derived
from using a unitary transformation method~\cite{sl} and  the
three-dimensional reductions~\cite{kl} of Bethe-Salpeter equation.
These formulations have been used in studying $\pi N$
scattering~\cite{pearce,sl,ntu}, $NN$ scattering~\cite{macheleidt},
and coupled-channel $\pi N$ and $\gamma N$ reactions in the nucleon
resonance region~\cite{ebac,ebac-1}.

In section II, we present the considered coupled-channel formulations
and discuss their differences with Eqs.(\ref{eq:vi-eq})-(\ref{eq:vi-g2}).
The numerical procedures for solving the considered coupled-channel
equations are described in section III.
We then investigate in section IV, the numerical consequences
of the differences between different coupled-channel models
in predicting the resonance positions and the reaction cross sections.
A summary is given in section V.

\section{Formalism}
\label{s2}

Following Ref.\cite{oset}, we assume that the interactions between the
considered meson-baryon ($MB$) channels are due to the
vector meson-exchange mechanism and can be calculated
 from
the following interaction Lagrangian
\begin{eqnarray}
{\cal L}_{int} = {\cal L}_{VVV} +{\cal L}_{PPV} +{\cal L}_{BBV}
\label{eq:lag}
\end{eqnarray}
with
\begin{eqnarray}
{\cal L}_{VVV}&=&ig\langle V^\mu[V^{\nu},\partial_\mu V_{\nu}]\rangle\nonumber\\
{\cal L}_{PPV}&=&-ig\langle V^\mu[P,\partial_\mu P]\rangle\nonumber\\
{\cal L}_{BBV}&=&g (\langle\bar{B}\gamma_\mu
[V^\mu,B]\rangle+\langle\bar{B}\gamma_\mu B\rangle\langle
V^\mu\rangle)\ \label{eq:lag-1}
\end{eqnarray}
where $P$ and $V$ stand for the Pseudoscalar and Vector mesons of the
16-plet of SU(4), respectively, and $B$ stands for the baryon.
The coupling constant
$g=M_V/2 f$ is taken from the hidden gauge model  with
 $f=93$ MeV being the pion
decay constant and $M_V=770$ MeV  the mass of the
light vector meson.

\begin{figure}[htbp] \vspace{-0.cm}
\begin{center}
\includegraphics[width=0.35\columnwidth]{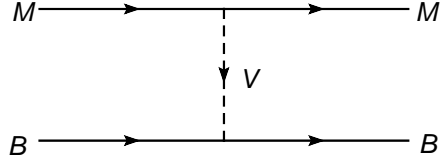}
\caption{One vector exchange mechanism of meson-baryon interactions.
 \label{Fey} }
\end{center}
\end{figure}

By using Eq.(\ref{eq:lag}), the invariant amplitude
of the $PB \rightarrow PB$ and $VB \rightarrow VB$ transitions
due to the one-vector-meson-exchange interaction, as illustrated
 as illustrated in Fig.\ref{Fey},
can be written as (suppressing the spin quantum numbers)
\begin{eqnarray}
{\cal M}^{PB,I,V}
(q_i,q_j)&=&C^{PB,I,V}_{i,j}\frac{M^2_{V}}{4f^2}
\frac{p^{\mu}_{V}p^{\nu}_{V}/m^2_{V}-g^{\mu\nu}}{p^2_{V}-m^2_{V}}
\bar{u}_{B_{i}}\gamma_{\mu}(p_{M_{i}}+p_{M_{j}})_{\nu}u_{B_{j}},
\label{pvf}\\
&& \nonumber \\
{\cal M}^{VB,I,V}
(q_i,q_j)&=&C^{VB,I,V}_{i,j}\frac{M^2_{V}}{4f^2}
\frac{p^{\mu}_{V}p^{\nu}_{V}/m^2_{V}-g^{\mu\nu}}{p^2_{V}-m^2_{V}}
 \bar{u}_{B_{i}}\gamma_{\mu}(p_{M_{i}}+p_{M_{j}})_{\nu}u_{B_{j}}
(-\varepsilon^*_{M_{i}}\cdot\varepsilon_{M_{j}}),\label{vvf}
\end{eqnarray}
where the sub-indices $i,j$ stand for the $M_iB_i$ and $M_jB_j$ channels,
$I$ is the total isospin of the system, $V$ denotes the
exchanged vector meson,
 $q_i$ is the relative momentum of the $ M_iB_i$
channel  in the center of mass frame, $p_\alpha$ is the
the four- momentum of particle $\alpha$,
$u_{B_i}$ is the Dirac
spinor of the  baryon $B_i$, and $\varepsilon_{M_i}$  is the
polarization vector of the external vector meson $M_i$.
The
coefficients $C^{PB, I,V}_{i,j}$ and $C^{VB, I,V}_{i,j}$ in Eqs.(\ref{pvf})
 and (\ref{vvf})
are taken from
Ref.\cite{oset} and listed in the
Tables \ref{coeff} and \ref{coeffv}.

\begin{table}[H]
 \renewcommand{\arraystretch}{1.2}
\centering

\caption{Coefficients $C^{PB,I,V}_{M_{i}B_{i} M_{j}B_{j}}$ in the
Eq.(\ref{pvf}) for the $PB$ system in the sector $I=1/2,3/2$,
$S=0$.  The
  exchanged vector mesons $V=\rho,\omega,D^*$ are indicated
next to the values of the coefficients.  } \label{coeff}
\begin{tabular}{l|c}
 I=3/2 & $\bar{D} \Sigma_{c}$  \\
 \hline
$\bar{D} \Sigma_{c}$ & $\rho$+$\omega$ \\
\end{tabular}
\\
\begin{tabular}{l|ccc}
 I=1 /2& $\bar{D} \Sigma_{c}$ & $\bar{D} \Lambda^{+}_{c}$ & $\eta_{c} N$  \\
 \hline
$\bar{D} \Sigma_{c}$     & -2$\rho$+$\omega$  &  $ 0$                      &  $-\sqrt{3/2}D^*$            \\
$\bar{D} \Lambda^{+}_{c}$&  $0$               &  $\omega$                  &  $\sqrt{3/2}D^*$              \\
$\eta_{c} N$             &  $-\sqrt{3/2}D^*$  &  $\sqrt{3/2}D^*$           &  $0 $                         \\
\end{tabular}
\end{table}

\begin{table}[H]
 \renewcommand{\arraystretch}{1.2}
\centering

\caption{Coefficients $C^{VB,I,V}_{M_{i}B_{i}\to M_{j}B_{j}}$ in the
Eq.(\ref{vvf}) for the $VB$ system in the sector $I=1/2,3/2$,
$S=0$.  The
  exchanged vector mesons $V=\rho,\omega,D^*$ are indicated
next to the values of the coefficients.} \label{coeffv}
\begin{tabular}{l|c}
 I=3/2 & $\bar{D}^* \Sigma_{c}$  \\
 \hline
$\bar{D}^* \Sigma_{c}$ & $\rho$+$\omega$ \\
\end{tabular}
\\
\begin{tabular}{l|cccc}
 I=1 /2& $\bar{D}^* \Sigma_{c}$ & $\bar{D}^* \Lambda^{+}_{c}$ & $J/\psi N$        &   $\rho N$\\
 \hline
$\bar{D}^* \Sigma_{c}$          & -2$\rho$+$\omega$  &  $ 0$                      &  $-\sqrt{3/2}D^*$     &  $-1/2 D^*$      \\
$\bar{D}^* \Lambda^{+}_{c}$     &  $0$               &  $\omega$                  &  $\sqrt{3/2}D^*$      &  $-3/2 D^*$       \\
$J/\psi N$                      &  $-\sqrt{3/2}D^*$  &  $\sqrt{3/2}D^*$           &  $0 $                 &  $0$     \\
$\rho N$                        &  $-1/2 D^*$        &  $-3/2 D^*$                &  $0 $                 &  $-2\rho$      \\
\end{tabular}
\end{table}

We consider the coupled-channel models derived by using the unitary transformation method of Ref.\cite{sl,Hyodopk}
and the three-dimensional reductions of Bethe-Salpeter equations employed in
 Ref.\cite{ntu}. In the center of mass (CM) frame,
the scattering equations within these models can be cast
into the following general form (suppressing the
 spin quantum numbers)
\begin{eqnarray}
\hat{T}^{\alpha,I}(\vec{q}_i, \vec{q}_j, \sqrt{s})&=&
\hat{V}^{\alpha,I}(\vec{q}_i, \vec{q}_j, \sqrt{s}) \nonumber \\
&&+ \sum_{k} \int d\vec{q}_k \hat{V}^{\alpha,I}(\vec{q}_i, \vec{q}_k, \sqrt{s})
\frac{N(\vec{q}_k,\sqrt{s})}{\sqrt{s}-E_{M_k}(\vec{q}_k)-E_{B_k}(\vec{q}_k) + i\epsilon}
\hat{T}^{\alpha,I}(\vec{q}_k, \vec{q}_j, \sqrt{s}) \nonumber \\
&&  \label{eq:3d-cceq}
\end{eqnarray}
where $\alpha = PB, VB$,
$\vec{q}_i$ is the relative three-momentum in channel $i$, $\sqrt{s}$ is the total energy,
and $E_{\alpha_i}(\vec{q}_i) = \sqrt{m^2_{\alpha_i} + \vec{q}^{\,\,2}_i}$
is the energy of
the particle $\alpha = M, B$  with a mass $m_{\alpha_i}$.
All external particles in the $MB$ channels are on their mass-shell.
Explicitly, we choose
\begin{eqnarray}
\vec{q}_i&=&\vec{p}_{M_i} = -\vec{p}_{B_i} \,,\nonumber \\
p_{M_i} &=&(E_{M_i}(\vec{p}_{M_i}), \vec{p}_{M_i})\,, \nonumber \\
p_{B_i} &=&(E_{B_i}(\vec{p}_{B_i}), \vec{p}_{B_i})
\label{eq:kine}
\end{eqnarray}

In Eq.(\ref{eq:3d-cceq}), the calculation of the driving term
$\hat{V}^{\alpha,I}(\vec{q}_i, \vec{q}_j, \sqrt{s})$ from the
invariant amplitude ${\cal M}^{\alpha,I,V}(q_i,q_j)$ of
Eqs.(\ref{pvf})-Eq.(\ref{vvf}) and the function
$N(\vec{q}_k,\sqrt{s})$ in the propagator depend on the
approximations used in deriving the above three-dimensional
equations from the relativistic quantum field theory. In the
following, we specify these two ingredients in each model.

\subsection{Model based on unitary transformation method}
\begin{enumerate}
\item $N(\vec{q}_k,\sqrt{s})=1$
\item The driving term is calculated from
the invariant amplitudes ${\cal M}$ of Eqs.(\ref{pvf})-(\ref{vvf}) by
expressing the four-momentum of
the exchanged vector meson in terms  of
the incloming and outgoing momenta. Explicitly, we use
Eq.(\ref{pvf}) for $PB \rightarrow PB$ to get
\begin{eqnarray}
&&\hat{V}^{PB,I,V}
(\vec{q}_i,\vec{q}_j) \nonumber \\
&&= C^{PB,I,V}_{i,j}
\frac{M^2_{V}}{4f^2}\bar{u}_{B_j}[\gamma_\mu({p}_{M_i}
+{p}_{M_j})_\nu] u_{B_i} \nonumber \\
&& \times \frac{1}{2}(\frac{\frac{(p_{M_i}-p_{M_j})^{\mu}(p_{M_i}-p_{M_j})^{\nu}}
{m^2_{V}}-g^{\mu\nu}}{(p_{M_i}-p_{M_j})^2-m^2_{V}}
+\frac{\frac{(p_{B_j}-p_{B_i})^{\mu}(p_{B_j}-p_{B_i})^{\nu}}{m^2_{V}}-g^{\mu\nu}}
{(p_{B_j}-p_{B_i})^2-m^2_{V}})
,\label{vpgt}
\end{eqnarray}
where the kinematic variables are given in Eq.(\ref{eq:kine}).

The procedure for calculating  the  $VB \rightarrow VB$ from Eq.(\ref{vvf})
is same since the factor
$-\epsilon^*_{M_i}\cdot \epsilon_{M_j}$ does not depend on $p_V$ of the
exchanged vector meson.
\end{enumerate}

\subsection{Models based on three-dimensional reductions}
We consider the three-dimensional reductions developed by Kadyshevsky, Blankenbecler and Sugar, and
Thompson,as explained in Ref.\cite{ntu}. All have the same form of the potential which
is defined by setting the time component of the
 four-momentum of the exchanged vector meson $V$ to  zero.
By using
Eq.(\ref{pvf}) for $PB \rightarrow PB$, we  get
\begin{eqnarray}
&& \hat{V}^{PB,I,V}
(q_i,q_j) \nonumber \\
&& = C^{PB,I,V}_{i,j}\frac{M^2_{V}}{4f^2} \frac{\bar{u}_{B_j}[
\frac{(\vec{p}_V\cdot\vec{\gamma})
(\vec{p}_V\cdot (\vec{p}_{M_i}+\vec{p}_{M_j}))}{M^2_V}
-(E_{M_i}(\vec{p}_{M_i})+E_{M_2}(\vec{p}_{M_j}))\gamma^0
 -  (\vec{p}_{M_i}+\vec{p}_{M_j})\cdot\vec{\gamma}]u_{B_i}}
{-\vec{p}_V^{\,\,2}-m^2_V}  \nonumber \\
&& \label{vpgt-a}
\end{eqnarray}
The procedure for calculating  the  $VB \rightarrow VB$
 from Eq.(\ref{vvf}) is same since the factor
$-\epsilon^*_{M_i}\cdot \epsilon_{M_j}$ doe not depend on $p_V$ of the
exchanged vector meson.

The function $N(\vec{k},\sqrt{s})$ of the propagator of Eq.(\ref{eq:3d-cceq}) for each reduction is
\begin{enumerate}
\item Kadyshevsky:
\begin{eqnarray}
N(q_i,\sqrt{s})=1
\end{eqnarray}

\item Blankenbecler-Sugar :
\begin{eqnarray}
N(q_i,\sqrt{s})={\frac{2(E_{M_i}(q_i)+E_{B_i}(q_i))}
{\sqrt{s}+(E_{M_i}(q_i)+E_{B_i}(q_i))}}
\end{eqnarray}
\item Thompson:
\begin{eqnarray}
N(q_i,\sqrt{s})={\frac{(E_{M_i}(q_i)+E_{B_i}(q_i))}{\sqrt{s}}}
\end{eqnarray}

\end{enumerate}

Note that for the on-shell momentum,
 defined by $\sqrt{s}= E_{M_i}(q_{0i}) + E_{B_i}(q_{0i})$,
 $N(q_{0i},\sqrt{s})=1$.
Thus all models satisfy the
same unitarity condition defined by the cuts of the propagators of
Eq.(\ref{eq:3d-cceq}).

\section{Calculation procedures}
In this section, we  describe our procedures for
solving the coupled-channel equations to obtain the
 $MB \rightarrow MB$ cross sections.

It is convenient to cast Eq.(\ref{eq:3d-cceq}) into the following familiar form
\begin{eqnarray}
{T}^{\alpha,I}(\vec{q}_i, \vec{q}_j, \sqrt{s})&=&V^{\alpha,I} (\vec{q}_i, \vec{q}_j, \sqrt{s}) \nonumber \\
&&+ \sum_{k} \int d\vec{q}_k{V}^{\alpha,I}(\vec{q}_i, \vec{q}_k, \sqrt{s})
\frac{1}{\sqrt{s}-E_{M_i}(\vec{q}_k)-E_{B_i}(\vec{q}_k) + i\epsilon}
{T}^{\alpha,I}(\vec{q}_k, \vec{q}_j, \sqrt{s}) \nonumber \\
\label{eq:3d-cceq-0}
\end{eqnarray}
where $\alpha = PB, VB$, and
\begin{eqnarray}
V^{\alpha,I} (\vec{q}_i, \vec{q}_j, \sqrt{s}) &=&
 N^{1/2}(q_i,\sqrt{s}) \sum_{V}\hat{V}^{\alpha,I,V}((\vec{q}_i, \vec{q}_j, \sqrt{s})
 N^{1/2}(q_j,\sqrt{s})
\label{eq:3d-v-0} \\
T^{\alpha,I}(\vec{q}_i, \vec{q}_j, \sqrt{s}) &=&
 N^{1/2}(q_i,\sqrt{s}) \hat{T}^{\alpha,I}((\vec{q}_i, \vec{q}_j, \sqrt{s})
 N^{1/2}(q_j,\sqrt{s})
\label{eq:3d-t-0}
\end{eqnarray}

With the normalization
$<\vec{p}|\vec{p}^{\,\,'}> = \delta(\vec{p}-\vec{p}^{\,\,'})$ for plane wave states,
we obtain from Eq.(\ref{eq:3d-cceq-0})
the following coupled-channel equations in each partial wave
\begin{eqnarray}
 T^{J,I}_{L_1,S_1,L_2,S_2}(q_1,q_2,\sqrt{s})&=&
V^{J,I}_{L_1,S_1,L_2,S_2}(q_1,q_2,\sqrt{s})\nonumber \\
&&+\sum_{L_3,S_3}\int q^2_3dq_3
 V^{J,I}_{L_1,S_1,L_3,S_3}(q_1,q_3,\sqrt{s})\ G(q_3,\sqrt{s})\ T^{J,I}_{L_3,S_3,L_2,S_2}(q_3,q_2,\sqrt{s})
\nonumber \\
&& \label{lt}
\end{eqnarray}
where $J$ is the total angular momentum; $L_i$ and $S_i$
are the orbital angular momentum and total spin of the $M_iB_i$ channel,
and the propagator is
\begin{eqnarray}
 G(q_i,\sqrt{s})&&=\frac{1}{\sqrt{s}-E_{M_i}(q_i)-E_{B_i}(q_i)}
\end{eqnarray}
The matrix elements of the potential in Eq.(\ref{lt})
 can be conveniently calculated from Eq.(\ref{eq:3d-v-0})
by using the LSJ-helicity transformation \cite{ebac}.
 Explicitly, we obtain
\begin{eqnarray}
 &&V^{J,I}_{L_1,S_1,L_2,S_2}(q_1,q_2,\sqrt{s})\nonumber \\
&& \ \ \ \ =F_{L_1,L_2}(q_1,q_2)
\frac{\sqrt{(2L_1+1)(2L_2+1)}}{2J+1}
\frac{1}{(2\pi)^3}\sqrt{\frac{m_{B_1}m_{B_2}}{2E_{M}(q_1)E_{B}(q_1)2E_{M}(q_2)E_{B}(q_2)}}
\nonumber\\
                                  &&\ \ \ \ \times   \sum_{V}
G^{I,V}_{1,2} \sum_{\lambda_{M_1}\lambda_{B_1}}
\sum_{\lambda_{M_2}\lambda_{B_2}} C^{J,M_{S_1}}_{L_1,S_1,0,M_{S_1}}
C^{S_1,M_{S_1}}_{j_{M_1}\lambda_{M_1},j_{B_1} -\lambda_{B_1}}
C^{J,M_{S_2}}_{L_2,S_2,0,M_{S_2}}
C^{S_2,M_{S_2}}_{j_{M_2}\lambda_{M_2},j_{B_2} -\lambda_{B_2}}\nonumber \\
                                  &&\ \ \ \ \times
N^{1/2}(q_1;\sqrt{s})<q_1;-\lambda_{B_1},\lambda_{M_1}|{\cal V}^J|
\lambda_{M_2},-\lambda_{B_2};q_2>
N^{1/2}(q_2,\sqrt{s}) \label{vlee}
\end{eqnarray}
where $\lambda_\alpha$ is the helicity of particle $\alpha$, and by
writting Eq.(\ref{vpgt}) or Eq. (\ref{vpgt-a}) (and also the similar forms for $VB$)
 in helicity representation we can evaluate
\begin{eqnarray}
&&<q_1;-\lambda_{B_1},\lambda_{M_1}|{\cal V}^J|\lambda_{M_2},-\lambda_{B_2};q_2>
\nonumber \\
&& \ \ \ \ \ \ =
 (2\pi)  \int^{1}_{-1}
d cos\theta\  d^{J}_{\lambda_{M_1}-\lambda_{B_1},\lambda_{M_2}-\lambda_{B_2}}(\theta)
\hat{V}^{PB/VB,I,V}_{\lambda_{M_1}\lambda_{B_1},\lambda_{M_2}\lambda_{B_2}}(q_1,q_2,\theta,\sqrt{s})
\nonumber \\
&& \label{vlee-1}
\end{eqnarray}
where $cos\theta = \hat{q}_1\cdot \hat{q}_2$, and the matrix element
in the integrand can be calculated by writing Eq.(\ref{vpgt}) or Eq.
(\ref{vpgt-a}) (and also the similar forms for $VB$) in helicity
representation for each of the considered coupled-channel models.

In Eq.(\ref{vlee}) $C^{J,M}_{j_1,m_{j_1},j_2,m_{j_2}}=<J,M|j_1,j_2,m_{j_1},m_{j_2}>$
is the  Clebsh-Gordon  coefficient,
the isospin factor is
\begin{eqnarray}
G^{I,V}_{1,2} = \sum_{m_{I_{M_1}}m_{I_{B_1}}}\sum_{m_{I_{M_2}}m_{I_{B
_2}}}C^{I,M_{I}}_{I_{M_1},I_{B_1},m_{I_{M_1}},m_{I_{B_1}}}
C^{I,M_{I}}_{I_{M_2},I_{B_2},m_{I_{M_2}},m_{I_{B_2}}} \,,
\end{eqnarray}
where $(I_{M}m_{I_{M}},
I_{B}m_{I_{B}})$ are the isospin quantum numbers of MB,
and the  form factor is chosen as
\begin{eqnarray}
F_{L_1,L_2}(q_1,q_2)=(\frac{\Lambda^2_{V}}{\Lambda^2_{V}+q_{1}^2})^{(\frac{L_1}{2}+2)}(\frac{\Lambda^2_{V}}{\Lambda^2_{V}+q_{2}^2})^{(\frac{L_2}{2}+2)},
\end{eqnarray}
where the cut-off parameter $\Lambda_{V}$ is assumed the same value
for all exchanged vector mesons for simplicity.

The differential cross sections are calculated from the partial-wave amplitudes by
\begin{eqnarray}
\frac{d\sigma}{d\Omega}&=&
\frac{16\pi^{4}}{s}\frac{q_2}{q_1}\frac{E_{M_1}E_{B_1}E_{M_2}E_{B_2}}
{(2j_{M_1}+1)(2j_{B_1}+1)}\sum_{m_{j_{M_1}}m_{j_{B_1}}}
\sum_{m_{j_{M_2}}m_{j_{B_2}}}|<M_2B_2|{T}(\sqrt{s})|M_1B_1>|^2,\label{cross}
\end{eqnarray}
with
\begin{eqnarray}
 <M_2B_2|&{T}&(\sqrt{s})|M_1B_1>\nonumber\\
       &=&<j_{M_2}m_{j_{M_2}}j_{B_2}m_{j_{B_2}},I_{M_2}m_{I_{M_2}}I_{B_2}m_{I_{B_2}}|{T}(\sqrt{s})|j_{M_1}m_{j_{M_1}}j_{B_1}m_{j_{B_1}},I_{M_1}m_{I_{M_1}}I_{B_1}m_{I_{B_1}}>\nonumber\\
       &=&\sum_{J,I}\sum_{L_1,S_1,L_2,S_2}T^{J,I}_{L_1,S_1,L_2,S_2}(q_1,q_2,\sqrt{s})Y_{L_2,M_{L_2}}(\theta, \phi) \sqrt{\frac{2L_1+1}{4\pi}}\nonumber\\
       &\times&C^{J,M_{J}}_{L_1,S_1,0,M_{S_1}}C^{S_1,M_{S_1}}_{j_{M_1},j_{B_1},m_{j_{M_1}},m_{j_{B_1}}}C^{J,M_{J}}_{L_2,S_2,M_{L_2},M_{S_2}}C^{S_2,M_{S_2}}_{j_{M_2},j_{B_2},m_{j_{M_2}},m_{j_{B_2}}}\nonumber\\
       &\times&C^{I,M_{I}}_{I_{M_1},I_{B_1},m_{I_{M_1}},m_{I_{B_1}}}C^{I,M_{I}}_{I_{M_2},I_{B_2},m_{I_{M_2}},m_{I_{B_2}}}.\label{T}
\end{eqnarray}
Obviously, $M_{J}=M_{S_1}=m_{j_{M_1}}+m_{j_{B_1}}$,
$M_{S_2}=m_{j_{M_2}}+m_{j_{B_2}}$,
$M_{L_2}=(m_{j_{M_1}}+m_{j_{B_1}})-(m_{j_{M_2}}+m_{j_{B_2}})$ and
$M_{I}=m_{I_{M_1}}+m_{I_{B_1}}=m_{I_{M_2}}+m_{I_{B_2}}$.

\section{The Results}
\label{s3}

In this section, we show the results from four models listed in
Sec.\ref{s2}, and then discuss their differences with previous
works~\cite{oset,wang,liuxiang}.

\subsection{The results of 4 models listed in Sec.\ref{s2}}

We first consider the model based on the unitary transformation
method described in subsection II.A. We determined the resonance
pole positions ($M_R= M-i\frac{\Gamma}{2}$) by using the analytic
continuation method of Ref.\cite{ssl}. We find that the resonance
positions are sensitive to the cutoff $\Lambda$, as seen in
Table~\ref{tpb12m}. The resonances are generated only when the
cutoff is larger than 800 MeV. As the cutoff $\Lambda$ changes from
800 MeV to 2000 MeV, the "binding energy" ($\Delta E = M- E_{thr}$)
is changed greatly from 0.002 MeV to 23.9 MeV. The corresponding
changes in imaginary parts are also very large.

These resonances are very close to the threshold of
$\bar{D}\Sigma_c$ in the PB sector and $\bar{D}^*\Sigma_c$ in the VB
sector. They are mainly caused by the strong attractive potential
from the t-channel $\rho$ meson exchange in $\bar{D}\Sigma_c \to
\bar{D}\Sigma_c$ and $\bar{D}^*\Sigma_c \to \bar{D}^*\Sigma_c$
processes. The situation here is different from the case when only
light flavors are involved.  For the $\pi N$ interaction, there is
no resonance below the $\pi N$ threshold, although the t-channel
$\rho$ meson exchange also provides attractive potential there with
a similar coupling constant. The differences between two cases are
mainly from the term $(p_{M_i}+p_{M_j})$ in Eq.({\ref{vpgt}}). The
potential is proportional to $(m_{M_i}+m_{M_j})$ near the threshold
of system. For the $\bar{D}\Sigma_c$ case, $(m_{M_i}+m_{M_j}) \sim
4$~GeV, while it is about $0.3$~GeV for the $\pi N$ case. Hence the
attractive potential of $\bar{D}\Sigma_c$ is an order of magnitude
stronger than that of $\pi N$. This give a natural explanation why
the PB system with heavy quarks can have quasi-bound states while
the corresponding pure light quark sector cannot. The similar thing
happens also for the VB system.

\begin{table}[H]
\renewcommand{\arraystretch}{1.2}
\caption{The pole position ($M-i\Gamma/2$) and ``binding energy"
($\Delta E=E_{thr}-M$) for different cut-off parameter $\Lambda$ and
spin-parity $J^P$. The threshold $E_{thr}$ is $4320.79$ MeV of
$\bar{D}\Sigma_c$ in PB system and $4462.18$ MeV of $\bar{D}^*
\Sigma_c$ in VB system. The unit for the listed numbers is MeV.}
\label{tpb12m} \centering
\begin{tabular}{cccccccc}
 \hline\hline
                            &           &    PB   System       &                & VB   System        &            \\
 \hline
  $J^{p}=\frac{1}{2}^{-}$   &$\Lambda$  &~~ $M-i\Gamma/2$      & ~~$\Delta E$   &~~ $M-i\Gamma/2$    &~~$\Delta E$ \\
 \hline
                            &$650$      &~~     -              & ~~  -          &~~      -           & ~~  -      \\
                            &$800$      &~~     -              & ~~  -          &~~$4462.178-0.002i$ & ~~  0.002  \\
                            &$1200$     &~~ $4318.964-0.362i$  & ~~ 1.826       &~~$4459.513-0.417i$ & ~~  2.667  \\
                            &$1500$     &~~ $4314.531-1.448i$  & ~~ 6.259       &~~$4454.088-1.662i$ & ~~  8.092  \\
                            &$2000$     &~~ $4301.115-5.835i$  & ~~ 19.68       &~~$4438.277-7.115i$ & ~~  23.90  \\
 \hline
 $J^{p}=\frac{3}{2}^{-}$    &           &                      & ~~             &                    &             \\
 \hline
                            &$650$      &~~     -              & ~~  -          &~~      -           & ~~  -      \\
                            &$800$      &~~     -              & ~~  -          &~~$4462.178-0.002i$ & ~~ 0.002  \\
                            &$1200$     &~~     -              & ~~  -          &~~$4459.507-0.420i$ & ~~ 2.673  \\
                            &$1500$     &~~     -              & ~~  -          &~~$4454.057-1.681i$ & ~~ 8.123  \\
                            &$2000$     &~~     -              & ~~  -          &~~$4438.039-7.268i$ & ~~ 23.14  \\
\hline\hline
\end{tabular}
\end{table}

\begin{figure}[htbp] \vspace{-0.cm}
\begin{center}
\includegraphics[width=0.5\columnwidth]{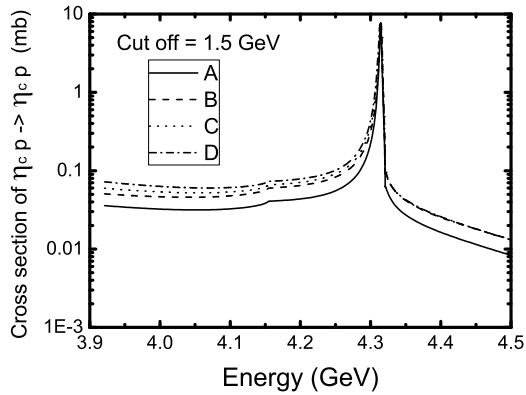}
\caption{The total cross section of $\eta_c p \to \eta_c p$ vs C.M.
energy is shown for 4 models. The 4 lines correspond to 4 models
listed in Table~\ref{modtab}.}\label{tcrs}
\end{center}
\end{figure}

The three other models based on three-dimensional reductions in
subsection II.B give similar results as shown in Table~\ref{modtab},
together with those from the model based on unitary transformation
method, taking the cut-off parameter $\Lambda=1500$ MeV. The
corresponding results for the total cross section of $\eta_c p
\rightarrow \eta_c p$ are shown in Fig.\ref{tcrs}. All these four
models predict a resonance below the $\bar{D}\Sigma_c$ threshold.
The masses and widths of the resonances from these different models
are almost the same.

\begin{table}[H]
\renewcommand{\arraystretch}{1.2}
\caption{Comparison for 4 models with the cut-off $\Lambda=1500$ MeV
and $J^P=1/2^-$ for PB system, where the threshold energy $E_{thr}$
is $4320.79$ MeV of $\bar{D}\Sigma_c$. ``A" is for the model based
on unitary transformation method; ``B" is for Kadyshevsky model;
``C" is for Blankenbecler-Sugar model; ``D" is for Thompson model.
$\Delta E_A$ and $\Gamma_A$ are the binding energy and width for the case A.
The unit is MeV.}\label{modtab} \centering
\begin{tabular}{ccccc}
 \hline\hline
   Models    &~~ $M-i\Gamma/2$          & ~~$\Delta E$  & ~~$|\frac{\Delta E- \Delta E_A}{\Delta E_A}|$ &~~$|\frac{\Gamma-\Gamma_A}{\Gamma_A}|$ \\
 \hline
     A          &~~ $4314.531-1.448i$      &~~  6.259      & ~~ 0         &~~  0      \\
     B          &~~ $4314.983-1.737i$      &~~  5.807      & ~~ 7.222\%   &~~  19.96\%      \\
     C          &~~ $4314.436-1.879i$      &~~  6.354      & ~~ 1.518\%   &~~  29.77\%      \\
     D          &~~ $4314.824-2.041i$      &~~  6.966      & ~~ 11.30\%   &~~  40.95\%      \\
\hline\hline
\end{tabular}
\end{table}

\subsection{Comparison with previous works}

\begin{figure}[htbp] \vspace{-0.cm}
\begin{center}
\includegraphics[width=0.5\columnwidth]{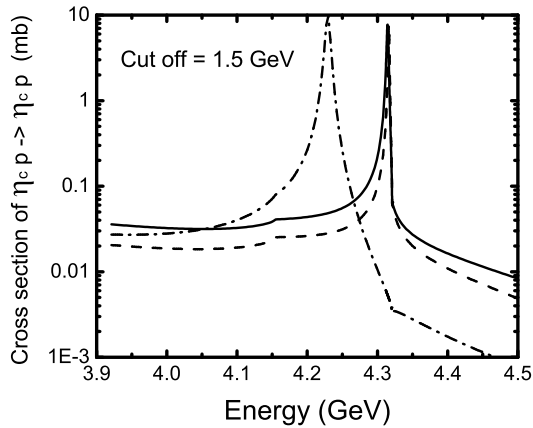}
\caption{The total cross section of $\eta_c p \to \eta_c p$ vs C.M.
energy for different potential approximations. The solid line is for
the full potential, corresponding to A in Table~\ref{osettab}; The
dashed line is for the neglect of spin of baryons, corresponding to
$A'$ in Table~\ref{osettab}; The dot-dashed line is for the neglect
of both spin of baryons and momentum of exchange vector,
corresponding to $A''$ in Table~\ref{osettab}.} \label{osetfi}
\end{center}
\end{figure}

In Ref.\cite{oset} using Valencia model, the mass and width of
predicted resonance in the PB system is about $4265$ MeV and $23$
MeV (for $\eta_c N$ channel only). Both binding energy and width are
much larger than the results in this work.  All models considered in
this work differ from the model used in Ref.\cite{oset} in
calculating the $MB \rightarrow MB$ potentials
Eqs.(\ref{pvf})-(\ref{vvf}). We will get the form used in
Ref.\cite{oset}, if we : (1) neglect the lower component of Dirac
spinor and keep only the time component $\gamma^0$; (2) set the
momentum squared of the exchanged vector meson $V$ to be
$p^2_V=(E^{on}_{M_i}-E^{on}_{M_j})^2-(q^{on}_i-q^{on}_j)^2 $ , where
the $E^{on}_{M_i}$ and $q^{on}_{i}$ are, respectively, the on-shell
energy and momentum of the meson in channel i. We have investigated 
the effects from taking each of these two assumptions. If we
only make the first simplification by neglecting spin of baryons,
then the corresponding results for the resonance are shown as for
potential $A'$ in Table~\ref{osettab}; If we continue to make the
second simplification by neglecting the momentum of the exchanged
vector meson, the results are shown as for potential $A''$ of
Table~\ref{osettab}. In Fig.\ref{osetfi}, we show the results
corresponding to these two simplifications (dashed line for $A'$ and
dot-dashed line for $A''$) for the $\eta_c p \rightarrow \eta_c p$
total cross section, compared with that (solid line for A) from the
model based on unitary transformation. Clearly, the second
simplification shifts the resonance position to a much lower value
and also increases the width significantly. This is the main reason
for the difference between our present results and those from
Ref.\cite{oset}. The second simplification makes $p^2_V$ and hence
the potential V independent on the integral momentum in
Eq.(\ref{eq:bs-eq}) so that Eq.(\ref{eq:bs-eq}) is simplified to
Eq.(\ref{eq:vi-eq}) instead of Eq.(\ref{eq:3d-cceq-0}) where the
potential V with integral momentum dependence is inside the
integration. Eq.(\ref{eq:vi-eq}) and Eq.(\ref{eq:3d-cceq-0}) give
the different results.

\begin{table}[H]
\renewcommand{\arraystretch}{1.2}
\caption{Comparison for different potential approximations with
cut-off $\Lambda=1500$ MeV and $J^P=1/2^-$ for PB system. The
threshold $E_{thr}$ is $4320.79$ MeV of $\bar{D}\Sigma_c$. ``A" is
for the full potential; ``$A'$" is for the neglect of spin of
baryons; ``$A''$" is for the neglect of both spin of baryons and
momentum of exchanged vector meson. $\Delta E_A$ and $\Gamma_A$ 
are the binding energy and width for the case A. The unit is MeV.}
\label{osettab} \centering
\begin{tabular}{ccccc}
 \hline\hline
   Potential    &~~ $M-i\Gamma/2$          & ~~$\Delta E$  & ~~$|\frac{\Delta E- \Delta E_A}{\Delta E_A}|$ &~~$|\frac{\Gamma-\Gamma_A}{\Gamma_A}|$ \\
 \hline
     A          &~~ $4314.531-1.448i$      &~~  6.259      & ~~ 0         &~~  0      \\
     $A'$       &~~ $4316.315-0.967i$      &~~  4.475      & ~~ 28.50\%   &~~  33.22\%      \\
     $A''$      &~~ $4229.362-3.914i$      &~~  91.43      & ~~ 1361\%    &~~  170.3\%      \\
\hline\hline
\end{tabular}
\end{table}

Besides Ref.\cite{oset}, there are two later
publications~\cite{wang,liuxiang} also predicting the existence of
$N^*$ around 4.3 GeV with hidden charm.

In Ref.\cite{wang}, the S-wave $\Sigma_c\bar D$ and $\Lambda_c\bar
D$ states with isospin I=1/2 and spin S=1/2 are dynamically
investigated within the framework of a chiral constituent quark
model by solving a resonating group method (RGM) equation. The
calculation not only includes vector mesons ($\rho$ and $\omega$)
exchange, but also scalar($\sigma$) meson exchange, which provides
an additional attractive force. Therefore, the binding energy in
Ref.\cite{wang} is larger than that in this work. The mass of the
bound state of $\bar{D} \Sigma_c$ is about $4279-4316$MeV.

In Ref.\cite{liuxiang}, the Schrodinger Equation was used to find
the bound state of $\bar{D} \Sigma_c$ and $\bar{D}^* \Sigma_c$ with
effective meson exchange potential. For the PB system, the $\rho$,
$\omega$ and $\sigma$ exchanges were considered. They tried the
different sign of coupling constants of various vertices. When they
chose the $\omega$ exchange to be repulsive, and the $\rho$ and
$\sigma$ exchange to be attractive, they also found the isospin 1/2
bound state of $\bar{D} \Sigma_c$ with cut off $\Lambda > 1.6$ GeV.
The binding energy is about $0-16$MeV corresponding to
$\Lambda=1.6-2.2$GeV, similar to the results in this work.

\section{Summary}

We have investigated the possible existence of nucleon resonances
with hidden charm within several coupled-channel models which are
derived from relativistic quantum field theory by using a unitary
transformation method and the  three-dimensional reductions of
Bethe-Salpeter Equation. With the same vector meson exchange
mechanism, we find that all models give very narrow molecular-like
nucleon resonances with hidden charm in the mass range of 4.3 GeV $
< M_R < $ 4.5 GeV, in consistent with the previous predictions. From
our analysis, the heavy mass of particles with the $c$ or $\bar{c}$
components would make the attractive potential stronger than the
case with only light flavors. The widths of these resonances are
very narrow in our models, because they need heavy vector meson
$D^*$ exchange to decay to open channels. Furthermore, we compare
our results with previous works. All of models predict a resonance
below the $\bar{D} \Sigma_c$ threshold. We also find that the pole
position would be shift a lot if we set
$p^2_V=(E^{on}_{M_i}-E^{on}_{M_j})^2-(q^{on}_i-q^{on}_j)^2$ for the
exchanged vector meson $V$ in the potential. We look forward to find
these predicted resonances with hidden charm in the reactions, such
as $e~p~\to~e~ J/\psi~p$, $p~p~\to~p~\eta_c(J/\psi)~p$, and
$p~\bar{p}~\to~p~\eta_c(J/\psi)~\bar{p}$.

Similarly the super-heavy $N^*$ with hidden beauty should also exist
although the binding energies may be not as large as given by the
simple Valencia model calculation of Ref.\cite{wubeauty}.

\bigskip
{\bf Acknowledgements:}

This work is Supported by the National Natural Science Foundation of
China (Nos. 10875133, 10821063, 11035006), the Chinese Academy
of Sciences Knowledge Innovation Project (Nos. KJCX2-EW-N01),
the Ministry of Science and Technology of China (2009CB825200), and
the U.S. Department of Energy, Office of Nuclear Physics Division,
under Contract No. DE-AC02-06CH11357.


\clearpage

\begin{thebibliography}{99}

\bibitem{PDG} Particle Data Group, C. Amsler {\it et al.},
Phys.\ Lett.\  {\bf B 667}, 1 (2008).

\bibitem{capstick} S. Capstick and W. Roberts,
Prog. Part. Nucl. Phys. \textbf{45}, S241 (2000).

\bibitem{Liubc} B.~C.~Liu, B.~S.~Zou, Phys. Rev. Lett. {\bf 96}, 042002 (2006);
ibid, {\bf 98}, 039102 (2007).

\bibitem{Weise} N.~Kaiser, P.~B.~Siegel and W.~Weise,
Phys.\ Lett.\  B {\bf 362}, 23 (1995).

\bibitem{or} E.~Oset and A.~Ramos,
  Nucl.\ Phys.\  A {\bf 635}, 99 (1998).

\bibitem{Osetoller} J.~A.~Oller, E.~Oset and A.~Ramos,
  Prog.\ Part.\ Nucl.\ Phys.\  {\bf 45}, 157 (2000).

\bibitem{meiss} J.~A.~Oller and U.~G.~Meissner,
  Phys.\ Lett.\  B {\bf 500}, 263 (2001).

\bibitem{Inoue} T.~Inoue, E.~Oset and M.~J.~Vicente Vacas,
  Phys.\ Rev.\  C {\bf 65}, 035204 (2002)

\bibitem{lutz}  C.~Garcia-Recio, M.~F.~M.~Lutz and J.~Nieves,
  Phys.\ Lett.\  B {\bf 582}, 49 (2004).

\bibitem{Hyodopk}
  T.~Hyodo, S.~I.~Nam, D.~Jido and A.~Hosaka,
  Phys.\ Rev.\  C {\bf 68}, 018201 (2003)

\bibitem{Riska} C.~Helminen and D.~O.~Riska,
Nucl.\ Phys.\  A {\bf 699}, 624 (2002).


\bibitem{Zou10} B.~S.~Zou,
  Nucl.\ Phys.\  A {\bf 835}, 199 (2010).

\bibitem{brodsky} S. J. Brodsky, P. Hoyer, C. Peterson and N. Sakat,
Phys.\ Lett.\ {\bf B 93}, 451 (1980).


\bibitem{oset}
  J.~-J.~Wu, R.~Molina, E.~Oset, B.~S.~Zou,
  Phys.\ Rev.\ Lett.\  {\bf 105}, 232001 (2010). Phys.\ Rev.\  {\bf C84}, 015202 (2011).

\bibitem{wang}
  W.~L.~Wang, F.~Huang, Z.~Y.~Zhang, B.~S.~Zou,
  Phys.\ Rev.\  {\bf C84}, 015203 (2011).

\bibitem{liuxiang}
  Z.~-C.~Yang, Zhifeng Sun, J.~He, X.~Liu, S.~-L.~Zhu,
  [arXiv:1105.2901 [hep-ph]].

\bibitem{afnan}
A.D. Lahiff and I.R. Afnan, PHys. Rev. C{\bf 66},044001 (2002)


\bibitem{sl}
T. Sato and T.-S. H. Lee, Phys. Rev. C{\bf 54}, 2660 (1996).

\bibitem{kl}
As reviewed by A. Klein and T.-S. H. Lee, Phys. Rev. D{\bf 10}, 4308 (1974)


\bibitem{pearce}
B.C. Pearce and B.K. Jennings, Nucl. Phys. {\bf 528}. 655 (1991)

\bibitem{ntu}
T. Hung, S.N. Yang, and T.-S. H. Lee,  Phys. Rev. C{\bf 64}, 034309 (2001)


\bibitem{macheleidt}
R. Macheleidt, Adv. Nucl. Phys. {\bf 19}, (1979)

\bibitem{ebac}
  A.~Matsuyama, T.~Sato and T.~S.~Lee,
  Phys.\ Rept.\  {\bf 439}, 193 (2007)
  [arXiv:nucl-th/0608051].


\bibitem{ebac-1}
 B.~Julia-Diaz, T.~-S.~H.~Lee, A.~Matsuyama, T.~Sato,
  Phys.\ Rev.\  {\bf C76}, 065201 (2007).
  [arXiv:0704.1615 [nucl-th]].


\bibitem{ssl}
N. Suzuki, T. Sato, and T.-S. H. Lee, Phys. Rev. C{\bf 79}. 025205 (2009)

\bibitem{wubeauty}
  J.~J.~Wu, L.~Zhao, and B.~S.~Zou,
  arXiv:1011.5743 [hep-ph], Phys. Lett. B (2012),
  doi:10.1016/j.physletb.2012.01.068.


\end{thebibliography}
\end{document}